# Inter-Cell Antenna Calibration for Coherent Joint Transmission in TDD System

Hanqing Xu, Yajun Zhao, Linmei Mo, Chen Huang, and Baoyu Sun
Baseband Algorithm Department
ZTE Corporation
Beijing, P.R.China
xu.hanqing@zte.com.cn

*Abstract*—**In this work the modeling and calibration method of reciprocity error in a coherent TDD coordinated multi-point (CoMP) joint transmission (JT) system are addressed. The modeling includes parameters such as amplitude gains and phase differences of RF chains between the eNBs. The calibration method used for inter-cell antenna calibration is based on precoding matrix indicator (PMI) feedback by UE. Furthermore, we provide some simulation results for evaluating the performance of the calibration method in different cases such as varying estimation-period, cell-specific reference signals (CRS) ports configuration, signal to noise ratio (SNR), phase difference, etc. The main conclusion is that the proposed method for inter-cell antenna calibration has good performance for estimating the residual phase difference.**

I. INTRODUCTION

For LTE TDD system, since the downlink (DL) and uplink (UL) transmissions share the same spectrum, the channel reciprocity allows DL channels to be estimated through the sounding reference signals (SRS) transmissions on the UL. However, channel reciprocity is guaranteed only for propagation channels. Since different RF chains are used in reception and transmission for each antenna and the properties of RF chains can be varied by temperature, humidity, etc., antenna calibration is necessary to fully exploit channel reciprocity. For single cell, intra-cell calibration has been discussed and implemented in LTE TDD system [1-7].

Coordinated multi-point (CoMP) transmission and reception is considered for LTE-Advanced Rel.11 as a tool to improve the coverage of high data rates, the cell-edge throughput, and also to increase system throughput [8]. The downlink CoMP can be categorized into joint transmission (JT), coordinated scheduling/beamforming (CS/CB), dynamic cell selection (DCS) and dynamic point blanking (DPB). Evaluations show that CoMP can offer performance benefits, especially joint transmission. As data to a single UE is simultaneously transmitted from multiple transmission points for coherent joint transmission, inter-cell antenna calibration should be performed, otherwise the JT performance will be degraded.

[10] provides two antenna array calibration schemes for cells configured in TDD CoMP: self-calibration and over-the-air calibration. In both schemes, the reciprocity of one TRX in one cell is selected as a global calibration reference. Then all other TRXs of all cells in the cooperating set must be aligned to it. Also, [11] discusses inter-cell antenna calibration for CoMP JT, and gives the performance evaluation result of inter-cell antenna calibration.

This paper introduces a procedure that is used to calibrate inter-cell antenna reciprocity. This procedure is based on PMI feedback by UE, which are suitable for LTE TDD CoMP JT system. Furthermore, the paper provides some simulation results for evaluating the performance of the procedure in different cases such as varying estimation-period, CRS ports configuration, SNR, phase difference, etc. It should be noted that here we mainly focus on the phase differences, while the amplitude differences are not treated in this work. Besides, this paper does not distinguish between the eNB and cell.

The paper is organized as follows. Section II contains the mathematical model for inter-cell antenna calibration. Section III describes the method used to calibrate the inter-cell channel reciprocity error. Section V provides some simulation results illustrating the performance of the described calibration method and Section VI concludes the paper.

II. CHANNEL MODELING

We can assume the downlink propagation channel is reciprocal to the uplink propagation channel due to the uplink and downlink share the same frequency band in a LTE TDD system. Considering the gains of eNB RF chains and UE RF chains, the effective downlink channel can be described as follows

$$\widehat{\mathbf{H}}_{DL} = \mathbf{U}_{RX} \cdot \mathbf{H}_{DL} \cdot \mathbf{B}_{TX} \quad (1)$$

where $\mathbf{H}_{DL}$ is the downlink propagation channel, $\mathbf{U}_{RX}$ and $\mathbf{B}_{TX}$ are receiving RF link gain at UE side and transmitting RF link gain at eNB side respectively as shown in Fig. 1. The effective uplink channel can be described similarly as follows

$$\widehat{\mathbf{H}}_{UL} = \mathbf{B}_{RX} \cdot \mathbf{H}_{UL} \cdot \mathbf{U}_{TX} \quad (2)$$

where $\mathbf{H}_{UL}$ is the uplink propagation channel, $\mathbf{B}_{RX}$ and $\mathbf{U}_{TX}$ are receiving RF link gain at eNB side and transmitting RF link gain at UE side respectively as shown in Fig. 1. According to the reciprocity between the downlink propagation channel and the uplink propagation channel, $\mathbf{H}_{UL} = (\mathbf{H}_{DL})^T$ can be obtained.

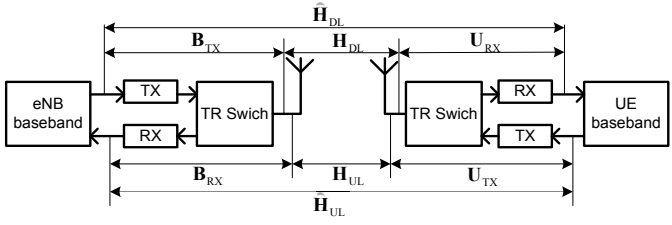

Figure 1. Channel reciprocity error modeling.

It was found that reciprocity error brought by UE RF link has little impact on the TDD system performance in [7, 11]. And so if not considering the impact of the antennas at UE,

$$\hat{\mathbf{H}}_{DL} = \mathbf{H}_{DL} \cdot \mathbf{B}_{TX} \quad (3)$$

$$\hat{\mathbf{H}}_{UL} = \mathbf{B}_{RX} \cdot \mathbf{H}_{UL} \quad (4)$$

then the downlink channel is given by

$$\begin{aligned}\hat{\mathbf{H}}_{DL} &= (\mathbf{H}_{UL})^T \cdot \mathbf{B}_{TX} \\ &= (\mathbf{B}_{RX}^{-1} \cdot \hat{\mathbf{H}}_{UL})^T \cdot \mathbf{B}_{TX} \\ &= (\hat{\mathbf{H}}_{UL})^T \cdot (\mathbf{B}_{RX}^{-1} \cdot \mathbf{B}_{TX})\end{aligned} \quad (5)$$

where $\mathbf{B}_{RX}^{-1} \cdot \mathbf{B}_{TX}$ denotes the eNB side reciprocity errors, and could be modeled as follows

$$\mathbf{C} = \mathbf{B}_{RX}^{-1} \cdot \mathbf{B}_{TX} = \begin{bmatrix} \frac{\alpha_{TX1}}{\alpha_{RX1}} & 0 & \cdots & 0 \\ 0 & \frac{\alpha_{TX2}}{\alpha_{RX2}} & \cdots & 0 \\ 0 & 0 & \ddots & 0 \\ 0 & 0 & \cdots & \frac{\alpha_{TXNt}}{\alpha_{RXNt}} \end{bmatrix} = \begin{bmatrix} a_1 e^{j\varphi_1} & 0 & 0 \\ 0 & \ldots & 0 \\ 0 & 0 & a_{Nt} e^{j\varphi_{Nt}} \end{bmatrix} \quad (6)$$

where Nt represents the antenna number of eNB. $a_i$ and $\varphi_i$ are the amplitude error and the phase error of antenna $i$ respectively, and

$$\mathbf{B}_{TX} = \begin{bmatrix} \alpha_{TX1} & \cdots & 0 \\ \vdots & \ddots & \vdots \\ 0 & \cdots & \alpha_{TXNt} \end{bmatrix}, \mathbf{B}_{RX} = \begin{bmatrix} \alpha_{RX1} & \cdots & 0 \\ \vdots & \ddots & \vdots \\ 0 & \cdots & \alpha_{RXNt} \end{bmatrix} \quad (7)$$

For single cell, inner-cell antenna calibration has been discussed in LTE TDD operation [1-7]. In this paper, we assume that the antennas in each cell have been already ideally calibrated, and two cells and one UE participate in the TDD CoMP joint transmission. The relation between the downlink channel and the uplink channel of each cell can be expressed as below before the calibration of the inter-cell antennas

$$\hat{\mathbf{H}}_{1,DL} = (\hat{\mathbf{H}}_{1,UL})^T \cdot \mathbf{C}_1, \quad \hat{\mathbf{H}}_{2,DL} = (\hat{\mathbf{H}}_{2,UL})^T \cdot \mathbf{C}_2 \quad (8)$$

where $\hat{\mathbf{H}}_{i,DL}$ ($i=1,2$), $\hat{\mathbf{H}}_{i,UL}$ ($i=1,2$) are the downlink channel matrix and the uplink channel matrix of cell $i$ in a CoMP JT set, respectively. $\mathbf{C}_1$, $\mathbf{C}_2$ are likely to be different after each cell calibrates its own antennas, due to not only the selected reference antenna but also the gains of RF chains between cells could be different. The combined downlink channel is given by

$$\hat{\mathbf{H}}_{DL} = \begin{bmatrix} \hat{\mathbf{H}}_{1,DL} + \hat{\mathbf{H}}_{2,DL} \end{bmatrix} \quad (9.a)$$

or

$$\hat{\mathbf{H}}_{DL} = \begin{bmatrix} \hat{\mathbf{H}}_{1,DL} & \hat{\mathbf{H}}_{2,DL} \end{bmatrix} \quad (9.b)$$

where $\hat{\mathbf{H}}_{DL}$ in (9.a) and (9.b) is the combined downlink channel of these two cells based on receiving downlink reference signals. If reference-signal ports of these two cells cannot be distinguished by UE, the combined channel should be given by (9.a). Otherwise, if reference-signal ports of two cells can be distinguished by UE, the combined channel should be given by (9.b). (9.a) and (9.b) can be rewritten as

$$\begin{aligned}\hat{\mathbf{H}}_{DL} &= (\hat{\mathbf{H}}_{1,UL})^T \cdot \mathbf{C}_1 + (\hat{\mathbf{H}}_{2,UL})^T \cdot \mathbf{C}_2 \\ &= (\hat{\mathbf{H}}_{1,UL})^T \cdot A_1 e^{j\theta_1} + (\hat{\mathbf{H}}_{2,UL})^T \cdot A_2 e^{j\theta_2} \\ &= \left( (\hat{\mathbf{H}}_{1,UL})^T + (\hat{\mathbf{H}}_{2,UL})^T \cdot \frac{A_2}{A_1} e^{j(\theta_2 - \theta_1)} \right) \cdot A_1 e^{j\theta_1}\end{aligned} \quad (10.a)$$

or

$$\begin{aligned}\hat{\mathbf{H}}_{DL} &= \begin{bmatrix} (\hat{\mathbf{H}}_{1,UL})^T \cdot \mathbf{C}_1 & (\hat{\mathbf{H}}_{2,UL})^T \cdot \mathbf{C}_2 \end{bmatrix} \\ &= \begin{bmatrix} (\hat{\mathbf{H}}_{1,UL})^T \cdot A_1 e^{j\theta_1} & (\hat{\mathbf{H}}_{2,UL})^T \cdot A_2 e^{j\theta_2} \end{bmatrix} \\ &= \begin{bmatrix} (\hat{\mathbf{H}}_{1,UL})^T & (\hat{\mathbf{H}}_{2,UL})^T \cdot \frac{A_2}{A_1} e^{j(\theta_2 - \theta_1)} \end{bmatrix} \cdot A_1 e^{j\theta_1}\end{aligned} \quad (10.b)$$

where $\mathbf{C}_1 = A_1 e^{j\theta_1}$, $\mathbf{C}_2 = A_2 e^{j\theta_2}$. Since coefficient $A_1 e^{j\theta_1}$ in (10.a) and (10.b) has no effect on the calculation of PMI, we can only consider the main contributor $\frac{A_2}{A_1} e^{j(\theta_2 - \theta_1)}$ to the reciprocity error of TDD CoMP JT.

### III. CALIBRATION FOR TDD COMP JT

The reciprocity error of TDD CoMP JT system is modeled by

$$\mathbf{C}_{JT} = \frac{A_2}{A_1} \cdot e^{j(\theta_2 - \theta_1)} \quad (11)$$

If we assume $\Delta A = \frac{A_2}{A_1}$ and $\Delta\theta = \theta_2 - \theta_1$, then the reciprocity error can be expressed as

$$\mathbf{C}_{JT} = \Delta A \cdot e^{j\Delta\theta} \quad (12)$$

Assuming maximal ratio transmission, the cells will use the beamforming weights after the inter-cell antenna calibration as

$$\mathbf{w}_1 = \frac{\left((\hat{\mathbf{H}}_{1,\text{UL}})^T\right)^H}{\left\|(\hat{\mathbf{H}}_{1,\text{UL}})^T\right\|}, \quad \mathbf{w}_2 = \frac{\left((\hat{\mathbf{H}}_{2,\text{UL}})^T\right)^H}{\left\|(\hat{\mathbf{H}}_{2,\text{UL}})^T\right\|} \cdot \mathbf{C}_{\text{JT}}^{-1} \quad (13)$$

Then the received signal at the UE side is given by

$$\begin{aligned}
\mathbf{r} &= \hat{\mathbf{H}}_{1,\text{DL}} \cdot \mathbf{w}_1 \cdot \mathbf{x} + \hat{\mathbf{H}}_{2,\text{DL}} \cdot \mathbf{w}_2 \cdot \mathbf{x} + \mathbf{n} \\
&= (\hat{\mathbf{H}}_{1,\text{UL}})^T \cdot \mathbf{C}_1 \cdot \frac{\left((\hat{\mathbf{H}}_{1,\text{UL}})^T\right)^H}{\left\|(\hat{\mathbf{H}}_{1,\text{UL}})^T\right\|} \cdot \mathbf{x} + \\
&\quad (\hat{\mathbf{H}}_{2,\text{UL}})^T \cdot \mathbf{C}_2 \cdot \frac{\left((\hat{\mathbf{H}}_{2,\text{UL}})^T\right)^H}{\left\|(\hat{\mathbf{H}}_{2,\text{UL}})^T\right\|} \cdot \mathbf{C}_{\text{JT}}^{-1} \cdot \mathbf{x} + \mathbf{n} \\
&= \mathbf{C}_1 \cdot \left(\sum_{i=1}^{2} \left\|(\hat{\mathbf{H}}_{i,\text{UL}})^T\right\|\right) \cdot \mathbf{x} + \mathbf{n}
\end{aligned} \quad (14)$$

where **r** and **x** are received signal and transmitted signal respectively, **n** denotes complex Gaussian noise vector. Obviously, the potential performance improvement can be achieved if we find an approach to estimate the reciprocity error $\mathbf{C}_{\text{JT}}$.

### A. Performance degradation caused by phase difference

Firstly, we evaluate the performance degradation caused by phase difference. Fig. 2 shows a). ideal transmission, i.e. there is none residual phase difference exists, and b). non-ideal transmission, i.e. there are residual phase differences $\pi/8$, $\pi/4$ and $\pi/2$, respectively. As shown in Fig. 2, trivial performance loss is introduced by phase difference $\pi/8$, and the degradation caused by phase difference $\pi/4$ is also acceptable. But when the phase difference is extend to $\pi/2$, obvious performance degradation will be noted. The worst case happened when the phase difference is uniformly distributed in $(-\pi, \pi]$, this is the case when any inter-cell antenna calibration is not treated. Consequently, phase difference calibration is necessary in a TDD CoMP-JT scenario.

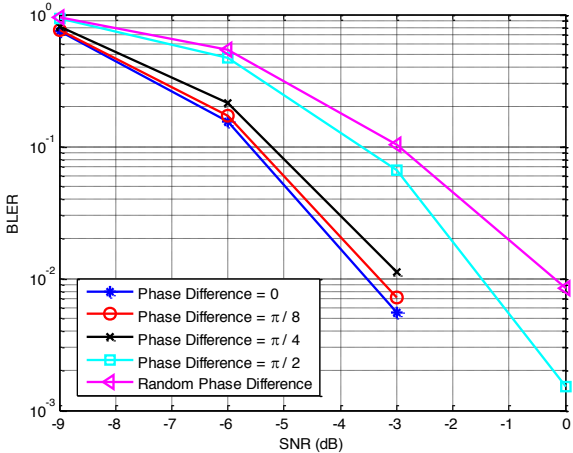

Figure 2. Performance degradation of different phase differences

### B. Estimation Algorithm of phase difference

Consider a CoMP-JT scenario with two transmission cells and one UE, and each cell is configured with one CRS port.

The combined downlink channel $\hat{\mathbf{H}}_{\text{DL}}$ received by UE has been expressed as (9.a) or (9.b) in section II. In fact, there is no difference between (9.a) and (9.b) for investigating inter-cell antenna calibration for CoMP JT. So the following of this paper takes CRS ports mapping scheme of (9.a) for example. As mentioned above, the phase difference information $\Delta\theta$ is contained in $\hat{\mathbf{H}}_{\text{DL}}$. The PMI calculated based on $\hat{\mathbf{H}}_{\text{DL}}$ at UE side can reflect the phase relationship between RF chains of these two eNBs, which means that different $\Delta\theta$ usually correspond to different PMI, and vice versa. Accordingly, $\forall \beta \in (-\pi, \pi]$, if $\beta$ is closer to $\Delta\theta$, the PMI calculated based on $\hat{\mathbf{H}}_{\text{DL}}^{\beta} = \left[(\hat{\mathbf{H}}_{1,\text{UL}})^T + (\hat{\mathbf{H}}_{2,\text{UL}})^T \cdot e^{j\beta}\right]$ at eNB side will be more similar to the PMI calculated based on $\hat{\mathbf{H}}_{\text{DL}}$ at UE side. Therefore, we can go through the phases $\beta_i \in (-\pi, \pi]$, compensate the uplink channel and calculate PMI based on it. If the PMI is similar or equal to the UE reported PMI, then the compensatory phase can be considered as an estimation of the actual phase difference. The detailed steps are as follows

*a)* By receiving the CRS transmitted by two cells, UE can estimates the combined downlink channel $\hat{\mathbf{H}}_{\text{DL}}$ given by (9.a) and (10.a).

*b)* The global precoding matrix indicator is obtained by $\hat{\mathbf{H}}_{\text{DL}}$ base on receiving CRS, and then UE may report a recommended $\mathbf{PMI}_{\text{UE}}$ feedback to two eNBs.

*c)* By receiving the SRS transmitted by UE, both eNB1 and eNB2 can estimate the uplink channel $\hat{\mathbf{H}}_{1,\text{UL}}$ and $\hat{\mathbf{H}}_{2,\text{UL}}$, respectively.

*d)* Assuming the existence of phase $\Delta\theta_1, \cdots, \Delta\theta_n$ ($\Delta\theta_i \in (-\pi, \pi]$), the second eNB will try to use $\Delta\theta_i$ to compensate residual phase difference between eNB1 and eNB2 as follows

$$\hat{\mathbf{H}}_{\text{DL}}^{\Delta\theta_i} = \left[(\hat{\mathbf{H}}_{1,\text{UL}})^T + (\hat{\mathbf{H}}_{2,\text{UL}})^T \cdot e^{j\Delta\theta_i}\right], \, i=1,\cdots,n \quad (15)$$

*e)* $\mathbf{PMI}_1, \cdots, \mathbf{PMI}_n$ can be calculated according to $\hat{\mathbf{H}}_{\text{DL}}^{\Delta\theta_i}$ in (15), respectively. If $\mathbf{PMI}_i$ fulfills the equation $\mathbf{PMI}_i = \mathbf{PMI}_{\text{UE}}$, it is clear that $\Delta\theta_i$ will be considered as the approximate value of $\Delta\theta$. Then we can implement the formula $f(\Delta\theta_i, m) = f(\Delta\theta_i, m-1) + 1$; Otherwise, we can implement the formula $f(\Delta\theta_i, m) = f(\Delta\theta_i, m-1)$, where $f(\Delta\theta_i, 0) = 0$, and $m$ denotes estimation sample index.

*f)* Repeat the process of step a) to e) until the end of the estimation period. Finally, the residual phase difference in this period can be calculated by

$$\Delta\hat{\theta} = \arg \max_{\substack{\Delta\theta_i \in (-\pi, \pi] \\ i=1,\cdots,n}} \left(f(\Delta\theta_i, m)\right) \quad (16)$$

## IV. SIMULATION RESULTS

In this section, we present the simulation performance of the approach provided in section III to estimate the residual phase difference. Without loss of generality, we assume that there are two transmission eNBs equipped with four inner-cell calibrated antennas respectively, and one UE equipped with two antennas in a CoMP JT set, as shown in Fig. 3. Fig. 3(a) shows the number of CRS ports configured in each eNB is four, and the ports can be distinguished at UE side are also four. The ports mapping mode is the same as (9.a) and (10.a) in section II. Fig. 3(b) shows the number of CRS ports configured in each eNB is two, and the ports can be distinguished at UE side are also two. The other simulation parameters are summarized in Table I.

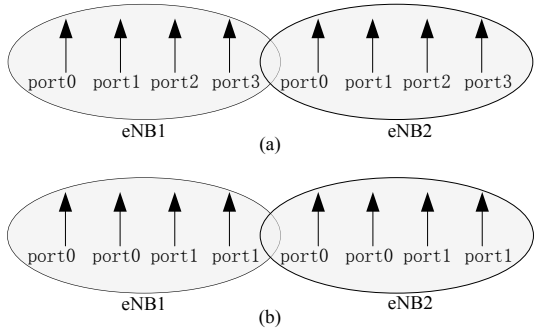

Figure 3.  (a) Configuration 1: four CRS ports configured in each eNB. (b) Configuration 2: two CRS ports configured in each eNB.

TABLE I.     SIMULATION PARAMETERS CONFIGURATION

| Scenario | Bandwidth | Propagation Channel | PMI Feedback Delay |
|---|---|---|---|
| TDD CoMP-JT | 10MHz | EPA 3 km/h | 10ms |

In the following we evaluate the performance of phase estimation algorithm for different cases such as varying estimation-period, CRS ports configuration, SNR, residual phase difference, etc. In all simulation cases, the residual phase differences are defined as multiples of a basic phase unit $\pi/8$, i.e. the residual phase difference belongs to the set $\{-7\pi/8, -6\pi/8, …, \pi\}$. Fig. 4 shows the results of simulation parameters including the actual residual phase difference $\Delta\theta = 6\pi/8$, SNR = 5dB, estimation-period = 10 frames, and four CRS ports configured in each eNB. From Fig. 5, we can see the estimation phase error is only $\pi/8$ which is in the acceptable range. In fact, we can use a longer estimation period to estimate the phase difference in the actual LTE TDD system.

### A. Estimation-period Results

Fig. 5 shows the simulation results of 1 frame estimation period. The actual residual phase difference $\Delta\theta$ is defined as $6\pi/8$ in both Fig. 4 and Fig. 5. The only difference between Fig. 4 and Fig. 5 is the estimation period, and other simulation parameters given in the figure caption and Table I are the same. The simulation results illustrate that compare to the 1 frame estimation period, the phase estimation algorithm that adopts 10 frames estimation period has a better performance. It can be concluded that the longer the estimation period, the more stable the estimated phase.

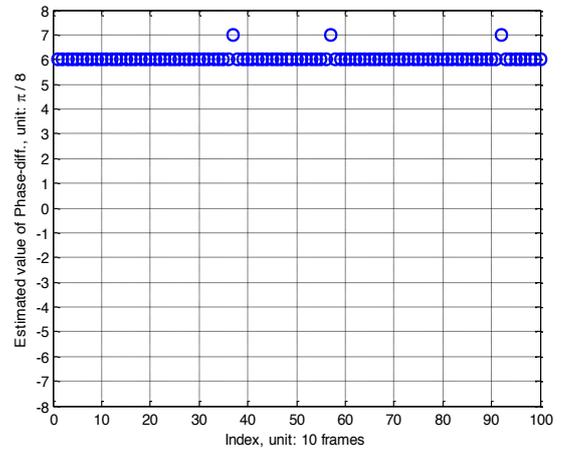

Figure 4.  the actual residual phase difference $\Delta\theta = 6\pi/8$, SNR = 5dB, estimation-period = 10 frames, four CRS ports are configured in each eNB.

### B. CRS Ports Configuration Results

As shown in Fig. 3, two CRS ports configurations of CoMP JT used for phase estimation algorithm simulation are given. Similarly, the actual residual phase difference $\Delta\theta$ is defined as $6\pi/8$ in both Fig. 4 and Fig. 6. The only difference between Fig. 4 and Fig. 6 is CRS ports configuration, and other simulation parameters given in the figure caption and Table I are the same. Fig. 4 and Fig. 6 illustrate the simulation results of different CRS port configurations, and the result of four CRS ports has a better performance.

### C. SNR Results

From Fig. 7, we can conclude a higher performance is achieved for a given SNR. Despite the fact that the performance is affected by SNR, the phase estimation algorithm has good performance if its estimation-period is ten frames or longer.

### D. Residual phase difference Results

In addition to the residual phase difference $6\pi/8$ in Fig. 4~ Fig. 7, we give the simulation results of another residual phase difference $\pi/8$ as shown in Fig. 8. From Fig. 8, we can observe the maximum estimation phase error is $\pm\pi/8$ which is also acceptable. The proposed phase estimation algorithm presents similar performance for different residual phase differences.

## V. CONCLUSIONS

In this paper, inter-cell reciprocity error was modeled and the model included parameters such as scale ratio and phase difference between the eNBs participating in the TDD CoMP JT. From this model, we observed that the performance of CoMP JT depends on the accuracy of inter-cell antenna calibration. If inter-cell antennas are not calibrated, SRS measurements cannot be used to calculate the downlink channels. Consequently, we presented a solution for inter-cell phase difference calibration in a TDD CoMP JT scenario. Furthermore, we provided the simulation results for evaluating the performance of our solution in different cases such as varying estimation-period, CRS ports configuration, SNR, residual phase difference, etc. Despite the fact that the performance is affected by port-configuration mode and SNR, the proposed phase estimation algorithm has good performance if its estimation-period is only ten frames. In fact, a longer estimation period is permitted in the actual TDD CoMP JT system.

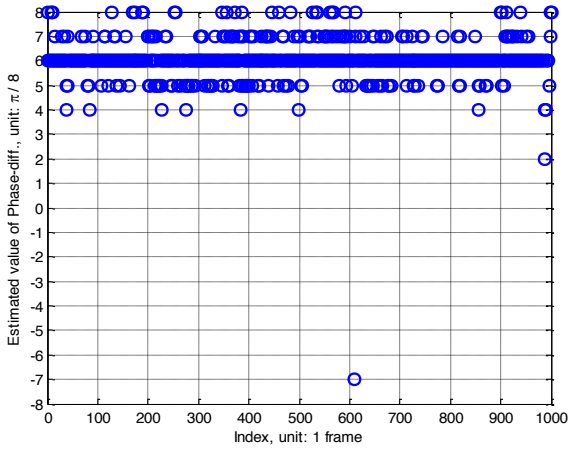

Figure 5. the actual residual phase difference $\Delta\theta = 6\pi/8$, SNR = 5dB, estimation-period = 1 frame, four CRS ports are configured in each eNB.

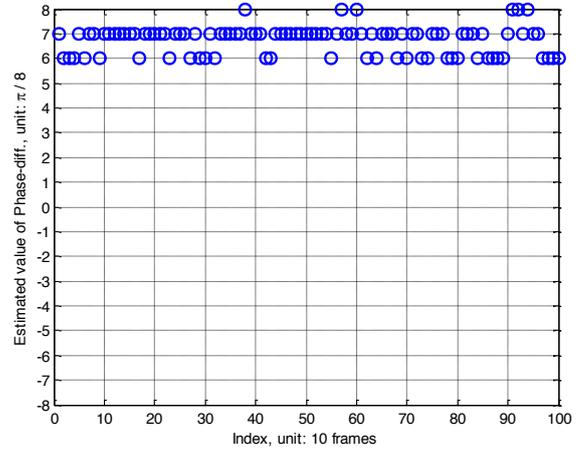

Figure 7. the actual residual phase difference $\Delta\theta = 6\pi/8$, SNR = −5dB, estimation-period = 10 frames, four CRS ports are configured in each eNB.

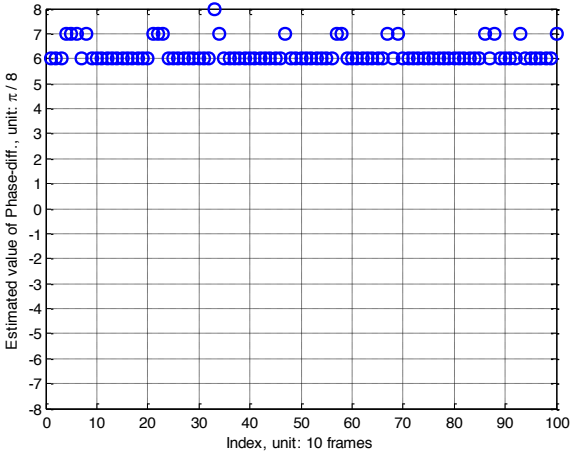

Figure 6. the actual residual phase difference $\Delta\theta = 6\pi/8$, SNR = 5dB, estimation-period = 10 frames, two CRS ports are configured in each eNB.

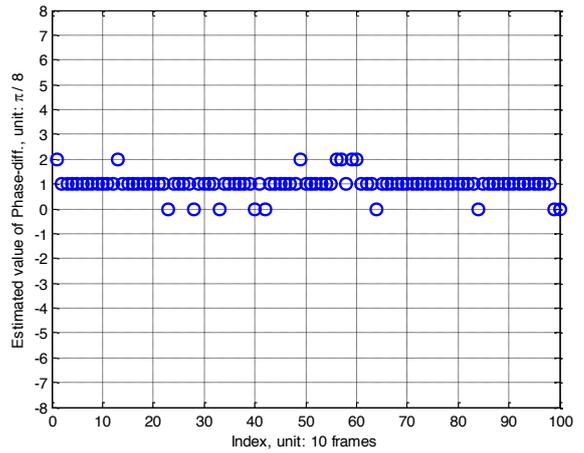

Figure 8. the actual residual phase difference $\Delta\theta = \pi/8$, SNR = 5dB, estimation-period = 10 frames, four CRS ports are configured in each eNB.

ACKNOWLEDGMENT

The authors would like to thank all their colleagues in the project at ZTE Corporation, especially Yujie Li, Jie Li, Peng Geng, Hongfeng Qin, Fan Shi, Wenzhong Zhang, Jun Huang.